\documentclass[conference]{IEEEtran}
\IEEEoverridecommandlockouts

\usepackage{cite}
\usepackage{amsmath,amssymb,amsfonts}
\usepackage{algorithmic}
\usepackage{graphicx}
\usepackage{textcomp}
\usepackage{xcolor}
\def\BibTeX{{\rm B\kern-.05em{\sc i\kern-.025em b}\kern-.08em
    T\kern-.1667em\lower.7ex\hbox{E}\kern-.125emX}}

\usepackage{hyperref}
\usepackage{booktabs}
\usepackage[section]{placeins}
\usepackage{multirow}
\usepackage{listings}
 
\colorlet{punct}{red!60!black}
\definecolor{background}{HTML}{EEEEEE}
\definecolor{delim}{RGB}{20,105,176}
\colorlet{numb}{magenta!60!black}

\lstdefinelanguage{json}{
    basicstyle=\normalfont\ttfamily,
    numbers=left,
    numberstyle=\scriptsize,
    stepnumber=1,
    numbersep=8pt,
    showstringspaces=false,
    breaklines=true,
    frame=lines,
    backgroundcolor=\color{background},
    literate=
     *{0}{{{\color{numb}0}}}{1}
      {1}{{{\color{numb}1}}}{1}
      {2}{{{\color{numb}2}}}{1}
      {3}{{{\color{numb}3}}}{1}
      {4}{{{\color{numb}4}}}{1}
      {5}{{{\color{numb}5}}}{1}
      {6}{{{\color{numb}6}}}{1}
      {7}{{{\color{numb}7}}}{1}
      {8}{{{\color{numb}8}}}{1}
      {9}{{{\color{numb}9}}}{1}
      {:}{{{\color{punct}{:}}}}{1} 
      {,}{{{\color{punct}{,}}}}{1} 
      {\{}{{{\color{delim}{\{}}}}{1} 
      {\}}{{{\color{delim}{\}}}}}{1} 
      {[}{{{\color{delim}{[}}}}{1} 
      {]}{{{\color{delim}{]}}}}{1}, 
}

\title{
Identification of the Most Frequently Asked Questions in Financial Analyst Reports to Automate Equity Research Using Llama 3 and GPT-4
}

\author{\IEEEauthorblockN{Adria Pop, Jan Spörer}
\IEEEauthorblockA{University of St.\,Gallen, Switzerland}}

\begin{document}\label{firstpage}




\maketitle

\begin{abstract}

This research dissects financial equity research reports (ERRs) by systematically mapping their content into categories. 
 


There is insufficient empirical analysis of the questions answered in ERRs. In particular, it is not understood how frequently certain information appears, what information is considered essential, and what information requires human judgment to distill into an ERR. 


The study analyzes 72 ERRs sentence-by-sentence, classifying their 4964 sentences into 169 unique question archetypes. We did not predefine the questions but derived them solely from the statements in the ERRs. This approach provides an unbiased view of the content of the observed ERRs. Subsequently, we used public corporate reports to classify the questions' potential for automation. Answers were labeled ``text-extractable'' if the answers to the question were accessible in corporate reports.






75.15\% of the questions in ERRs can be automated using text extraction from text sources. Those automatable questions consist of 51.91\% text-extractable (suited to processing by large language models, LLMs) and 24.24\% database-extractable questions. Only 24.85\% of questions require human judgment to answer.

We empirically validate, using Llama-3-70B and GPT-4-turbo-2024-04-09, that recent advances in language generation and information extraction enable the automation of approximately 80\% of the statements in ERRs. Surprisingly, the models complement each other's strengths and weaknesses well, indicating strong ensemble potential.

The research confirms that the current writing process of ERRs can likely benefit from additional automation, improving quality and efficiency. 
The research thus allows us to quantify the potential impacts of introducing large language models in the ERR writing process.

The full question list, including the archetypes and their frequency, are available online (\href{https://janspoerer.github.io/pop-spoerer-2025-financial-report-data}{janspoerer.github.io/pop-spoerer-2025-financial-report-data}).

\end{abstract}




\section{Introduction}\label{section_introduction_ffaq}


This study evaluates the automation potential of equity research reports by classifying analyst statements into categories and identifying which report components require human judgment.
One of the key contributions is a question list that analysts answer in ERRs. This question list
gives a holistic overview of ERR topics. We classify each question's automation potential by comparing the data sources to the statements to see whether an automated system could have written the ERRs.
If the share of automatable questions is high, this study may indicate that the automation of large parts of ERRs is feasible.




ERRs are written mainly by sell-side banks and specialized research companies. The purpose of ERRs is ultimately to serve as a buy, hold, or sell recommendation~\cite[p.\ 246]{AsquithMikhailAu_InformationContentOfEquityAnalystReports_2005}. 
Two of the most common differences among ERRs are their content and intention. The content of ERRs can be to initiate a company's coverage, provide an ordinary update to an already covered company, or provide an extraordinary update about a company. Subsequent recommendations to buy, hold, or sell may change or stay the same.
ERRs often follow a specific update frequency. One common update cycle for ERRs is quarterly ordinary updates. ERRs thus often contain recent facts from quarterly company-provided financial reports. Half of all newly issued ERRs fall into this category, being released closely after the company published new information~\cite[p.\ 247]{AsquithMikhailAu_InformationContentOfEquityAnalystReports_2005}. 

We review the literature on ERRs in section~\ref{section_literature_financial_economics_and_equity_research_reports_ffaq}, discussing their importance, accuracy, and existing approaches for automation. 

\begin{figure}
    \ \\
    \ \\ 
    \ \\
    \caption{%
    \textbf{Histogram of Category Frequency, Grouped by Question Category.}
    The histogram shows how frequently different information is displayed, grouped by question category. ERRs contain the \textit{Financials} category most frequently and usually display this category in text-or-tabular format or in text-tabular-or-graphical or tabular format. \textit{Company} and \textit{Market} information are commonly in text-only format. Notably, \textit{Stock} information is rarely in text-only format. Only 2.82\% of questions (weighted by their number of occurrences) were exclusively displayed in graphical format.}\label{fig:histogram_grouped_by_question_category_x_type_of_display_y_occurrences}
    \begin{center}
        \includegraphics[width=9.0cm]{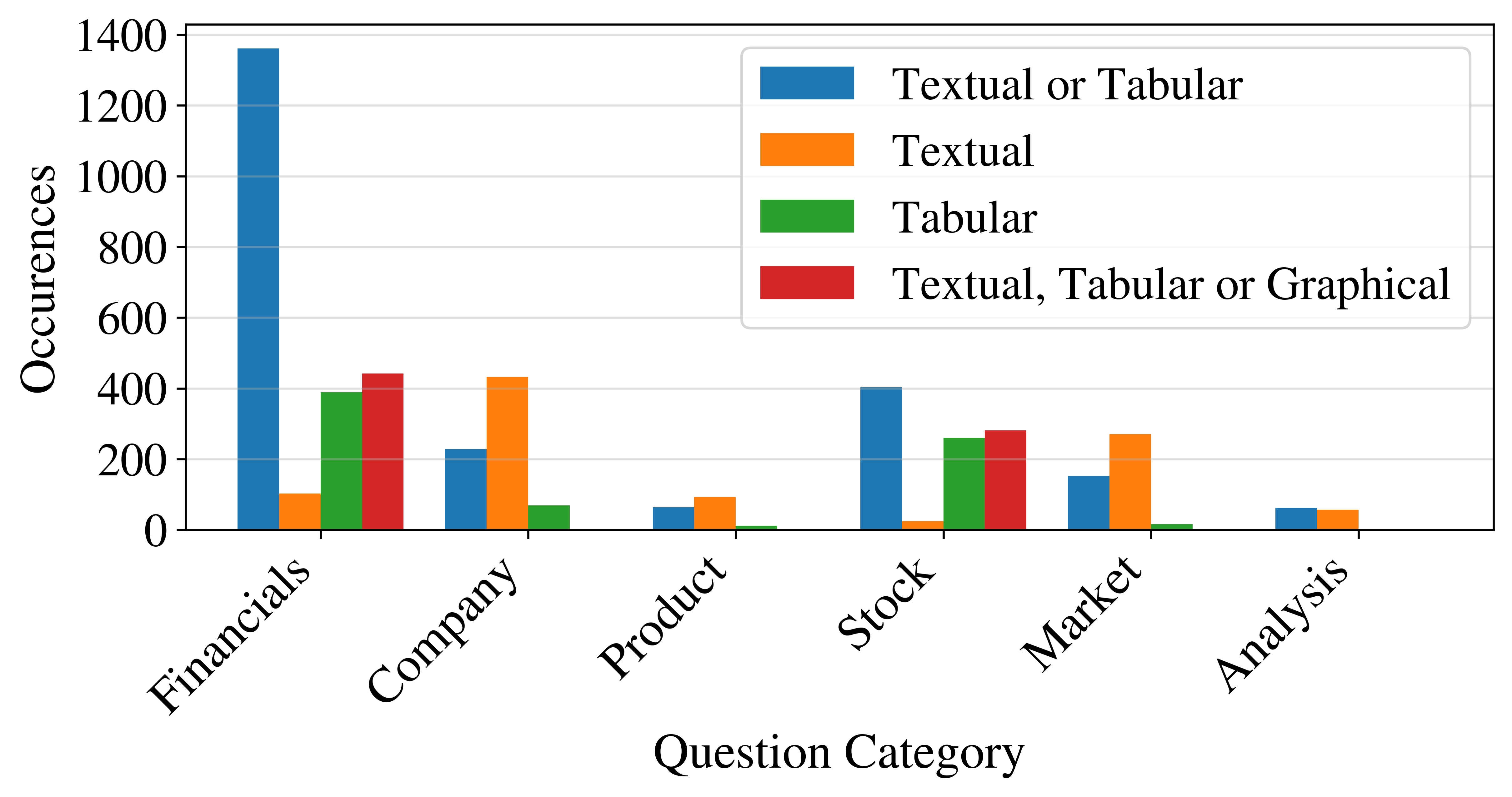}
    \end{center}
\end{figure}

We created the question list by manually reading each sentence of 72 ERRs. We mapped each phrase in the reports to a question. When we encountered a new question, we added it to the list. The result is a histogram of question occurrences (figure~\ref{fig:histogram_grouped_by_question_category_x_type_of_display_y_occurrences}). 
We provide more detail about the methodology in section~\ref{section_methods_ffaq}. 

This approach aligns with prior research \cite[p.\ 251]{AsquithMikhailAu_InformationContentOfEquityAnalystReports_2005}, with the main difference being that we did not assume specific data fields (or questions) to be present in the reports ex-ante. Instead, we recorded each statement, derived a question from it, and counted the number of occurrences of each question. This approach ensures maximum unbiasedness in representing the landscape of ERRs. 


No systematic reviews of ERR automation exist.
News feeds for financial news are already partially generated by AI systems. However, ERR writing is not yet widely automated, albeit being feasible~\cite{colemanMerkleyPacelli_HumanVersusMachineAComparisonOfRoboAnalystAndTraditionalResearchAnalystInvestmentRecommendations_2022}. Some consumer-grade analyst houses such as Zacks.com use template-based automation to update their company profiles and overview articles. 
Longer texts, however, are still written by humans. ERRs are one example of such longer texts, and they are the focus of this study. 


\section{Literature}\label{section_literature_ffaq}

\subsection{Review of the Financial Economics Literature}\label{section_literature_financial_economics_and_equity_research_reports_ffaq}


Considering how strongly stock market prices react after ERRs are published, their importance for investors and, by extension, stock-listed corporations is evident in the literature \cite{bjerringLakonishokVermaelen_StockPricesAndFinancialAnalystsRecommendations_1983,elton_GruberGrossman_DiscreteExectationalDataAndPortfolioPerformance_1986,liu1_SmithSyed_StockPriceReactionsToTheWallStreetJournalsSecuritiesRecommendations_1990,beneish_StockPricesAndTheDisseminationOfAnalystsRecommendation_1991,stickel_TheAnatomyOfThePerformanceOfBuyAndSellRecommendations_1995}.



\cite{Womack_DoBrokerageAnalystsRecommendationsHaveInvestmentValue_1996} observed that analysts could predict the directionality of stock returns six months into the future (pp.\ 139, 163--165), indicating significant information content in ERRs. At least directionally, not necessarily with respect to the accuracy of the price targets, ERRs thus exhibit predictive accuracy higher than what can be expected from random guessing.

A persistent problem with the reliability of ERRs is the reluctance of analysts to present negative recommendations, as \cite{barberLehavyMcnicholsTrueman_CanInvestorsProfitFromTheProphetsSecurityAnalystRecommendationsAndStockReturns_2001} and \cite{MikhailWaltherWillis_DoSecurityAnalystsExhibitPersistentDifferencesInStockPickingAbility_2004} demonstrated. \cite{AsquithMikhailAu_InformationContentOfEquityAnalystReports_2005} report a rate of sell recommendations of only 0.5\% in an ERR sample from 1997 to 1999 (p.\ 255), while \cite{Womack_DoBrokerageAnalystsRecommendationsHaveInvestmentValue_1996} report a rate of sell recommendations of 14\% (p.\ 164) between 1989 and 1991. Prior research by \cite{michaelyWomack_ConflictOfInterestAndTheCredibilityOfUnderwriterAnalystRecommendations_1999} concludes that the reason for the low number of sell recommendations is likely that the companies covered by many financial analysts are the banks' clients. Therefore, incentives arise to report too positively.

According to a study by \cite{AsquithMikhailAu_InformationContentOfEquityAnalystReports_2005}, more than half (54\%) of analyst reports set price targets that are achieved within a year (pp.\ 278--279). This accuracy rate appears relatively low, considering that the majority of ERRs tend to offer conservative recommendations, with price targets slightly exceeding current prices, on average~\cite[p.\ 256]{AsquithMikhailAu_InformationContentOfEquityAnalystReports_2005}.
The authors did not judge whether the analysts' success rate was good or poor. In contrast to this outcome, \cite{BradshawBrownHuang_DoSellSideAnalystsExhibitDifferentialTargetPriceForecastingAbility_2013} found that only 38\% of analysts' price targets are met within a one-year horizon (pp.\ 953--954). \cite{BoniniZanettiBianchiniSalvi_TargetPriceAccuracyInEquityResearch_2010} provide another critical analysis of financial analysts' accuracy~(pp.\ 1177, 1193--1196, 1208). The unequivocal findings potentially strengthen the argument for automating ERRs to achieve more robust price targets~\cite[pp.\ 80--81, for further discussion on report accuracy]{GleasonJohnsonLi_ValuationModelUseAndThePriceTargetPerformanceOfSellSideEquityAnalysts_2013}.





Research by~\cite{colemanMerkleyPacelli_HumanVersusMachineAComparisonOfRoboAnalystAndTraditionalResearchAnalystInvestmentRecommendations_2022} shows that these issues can be mitigated by using a hybrid machine-human approach. The study presents a computer-aided approach that better balances the buy, hold, and sell recommendation frequencies, achieves better portfolio performance, and reduces the time required for writing reports.

The previously presented studies show that the importance and accuracy of ERRs are studied extensively. Furthermore, there is extensive research on how stock performance is predicted by systematic factors, notably factor models by~\cite{FamaFrench_TheCrossSectionOfExpectedStockReturns_1992, FamaFrench_SizeAndBookToMarketFactorsInEarningsAndReturns_1995, FamaFrench_ChoosingFactors_2018, FamaFrench_AFiveFactorAssetPricingModel_2015} and by other studies from the field of risk factors and asset pricing~\cite{Carhart_OnPersistenceInMutualFundPerformance_1997, Asness_ValueAndMomentumEverywhere_2013}.\label{section_literature_review_automation_of_equity_research_ffaq}

\cite{AsquithMikhailAu_InformationContentOfEquityAnalystReports_2005} performed a similar empirical analysis as our study. They ex-ante determined 30 variables of interest that were extracted from ERRs. There are differences in the objective, the data, and the methodology when comparing \cite{AsquithMikhailAu_InformationContentOfEquityAnalystReports_2005} to our study. They examined the accuracy of ERRs, how much they impact markets, and how independent research providers compare to sell-side banks. The data is from 1997 to 1999, and they selected only a subset of high-performing analysts. Also, they derived the 30 data fields from the objective before reading the ERRs; in this study, however, we read the ERRs and generated the questions ad-hoc when encountering new questions. 

\begin{figure}
    \ \\
    \caption{\textbf{Histogram With Statement Counts in ERRs by Frequency.}  ERRs rarely have less than 30 statements. Most reports have between 30 and 119 statements.}\label{fig:statement_counts_histogram_binned}
    \begin{center}
        \includegraphics[width=6.5cm]{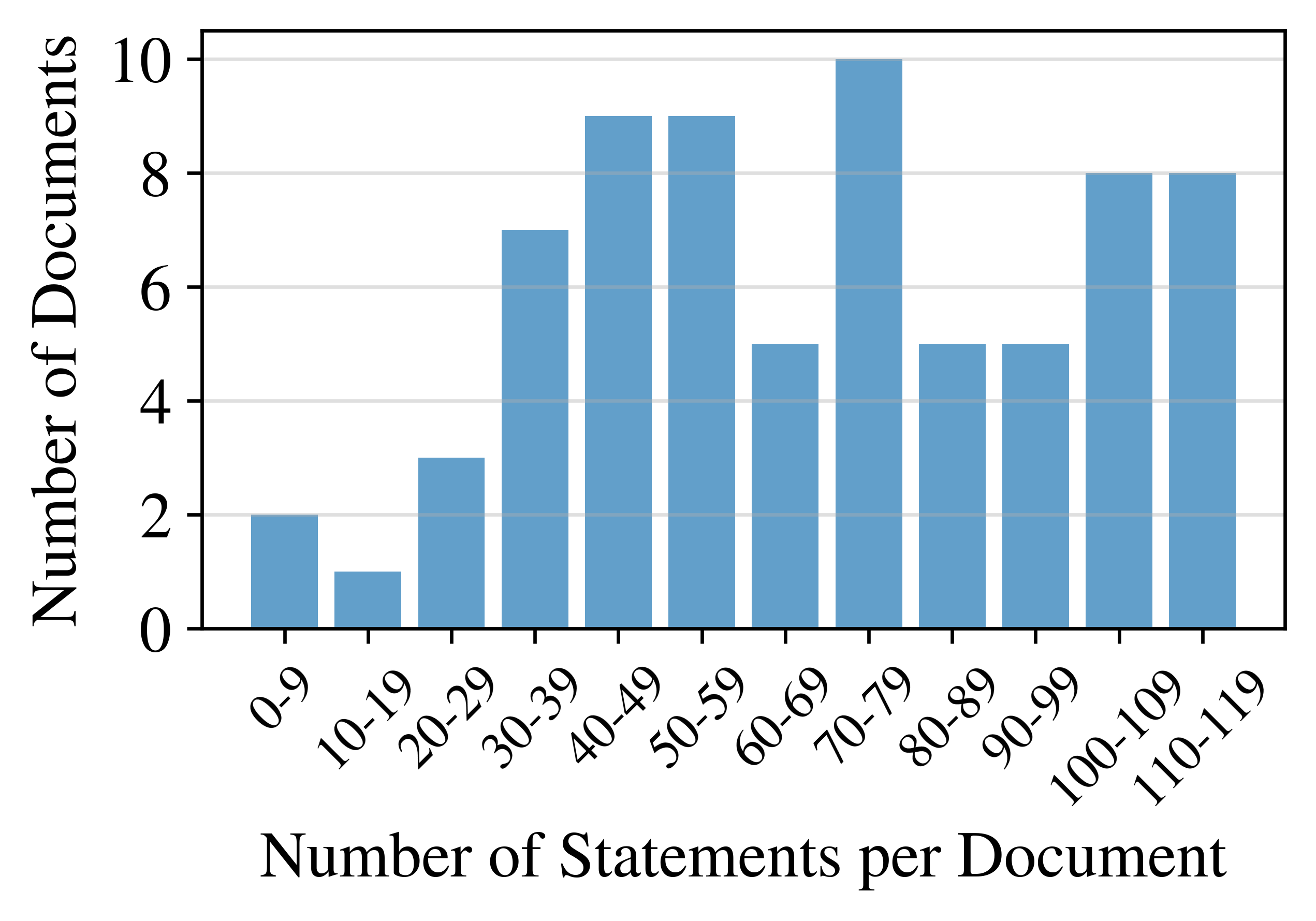}
    \end{center}
\end{figure}

\cite{kimMuhnNikolaev_FinancialStatementAnalysisWithLargeLanguageModels_2024} recently showed that the large language model \mbox{GPT-4} can slightly outperform a simpler neural network in processing equity risk factors for stock analysis. They also show that humans are significantly worse at predicting stock returns than the authors' language model-based system. This finding is in line with the aforementioned biases of human analysts. 


\subsection{Review of the Technical Literature}


Text retrievers get a question or a topic as input and search for fitting segments from a knowledge base. The retrieved span of text can be used directly as the answer or can be postprocessed by another system. 
Text retrieval from financial statements, even in combination with text from table captions, is already implemented by \cite{ChenChenSmileyShahBorovaLangdonMoussaBeaneHuangRoutledgeWang_FinQaADatasetOfNumericalReasoningOverFinancialData_2021}, and earlier research on business text extraction has been a field of interest for years \cite{thaiDavidORiainOSullivanHandschuh_SemanticallyEnhancedPassageRetrievalForBusinessAnalysisActivity_2008}. 
As retrievers are capable of choosing facts from knowledge bases of practically unlimited size, they are a key lower-level component needed to enable fact-based automated writing \cite{BorgeaudMenschHoffmann_DeepMind_RETRO_ImprovingLanguageModelsByRetrievingFromTrillionsOfTokens_2021}. It thus remains a core technology for automated ERR writing, at least until large language models' context sizes become larger than they are today~\cite{chenWongChenTian_ExtendingContextWindowOfLargeLanguageModelsViaPositionalInterpolation_2023}.


As the stated goal of this study is to pave the way for automating the writing of ERRs, we provide a brief overview of the question answering (QA) literature. QA is a subfield of NLP. It can be one of the domains that facilitate the automation of ERR writing.

Dense passage retrieval uses a latent representation of a question to search for an answer in a large corpus of text~\cite{karpukhinOguzMinLewisWuEdunovChenYih_DensePassageRetrievalForOpenDomainQuestionAnsewring_2020}. Combining such a retrieval mechanism with a generative language model by including the retrieval outputs to the language model prompt, one gets a retrieval-augmented generator (RAG) as presented by \cite{LewisPerezPiktusPetroniGoyalKuttlerLewisYihRocktaschelRiedelKiela_RetrievalAugmentedGenerationForKnowledgeIntensiveNlpTasks_2020}.
RAG continues to receive attention from the research community, as follow-up research on the topic shows~\cite{GuuLeeTungPasupatChang_REALM_RetrievalAugmentedLanguageModelPretraining_2020,BorgeaudMenschHoffmann_DeepMind_RETRO_ImprovingLanguageModelsByRetrievingFromTrillionsOfTokens_2021,izacardLewisLomeliHosseiniPetroniSchickDwivediyuJoulinRiedelGrave_Atlat_FewShotLearningWithRetrievalAugmentedLanguageModels}. Its effectiveness in writing factually correct and fluent text makes RAG a technology that may facilitate the automation of ERR writing.


With the emergence of the mentioned RAG methods, generative models are better capable of performing QA tasks. Furthermore, as models and their training data scale, their ability to store knowledge in their parameters increases, making them capable QA models even without external knowledge bases, as GPT-3~\cite{brownMannRyderSubbiahKaplanDhariwal_GTP3_LanguageModelsAreFewShotLearners_OpenAi_2020} and the Llama models~\cite{touvronLavrilIzacardMartinetLachausLacroixRoziereGoyalHambroAzharRodriguezJoulinGraveLample_Llama_OpenAndEfficientFoundationLanguageModels_2023} demonstrate. In addition, advances in the expansion of the context length \cite{PressSmithLewis_TrainShortTestLongAttentionWithLinearBiasesEnablesInputLengthExtrapolation_2022, chenWongChenTian_ExtendingContextWindowOfLargeLanguageModelsViaPositionalInterpolation_2023} make it possible to provide more world knowledge into prompts.

As the general capabilities of language models grow, finance-specific language models also improve. \cite{WuIsroyLuDabravolskiDredzeGehrmannKambadurRosenbergMann_BloombergGPT_ALargeLanguageModelForFinance_2023} develop a language model capable of understanding the nuances of a financial text. About half of their training data is finance-specific, and model performs well on financial QA (pp.\ 31--32).
These generative language model developments increase the capabilities of state-of-the-art language models to generate factually correct ERRs.




\section{Data}\label{section_data_ffaq}

We downloaded 72 ERRs dated from 2018 to 2023 from Bloomberg and Refinitiv Eikon. Each report had an average of seven pages, and the median is 6.8 pages per report. The shortest report is a one-pager, and the longest report has 20 pages. 
We analyzed 493 pages across all reports. These statistics are in line with findings of previous research by \cite[p.\ 252]{AsquithMikhailAu_InformationContentOfEquityAnalystReports_2005}. We sourced the ERRs from 23 different research providers, each contributing between 1 and 16 reports. 

The number of statements per report ranges from nine to 115. The average is 68, and the median is 69. See also figure~\ref{fig:statement_counts_histogram_binned} for a histogram of statement counts. In sum, we analyzed 4964 statements (sentences).

\section{Methods}\label{section_methods_ffaq}

\begin{figure}
    \ \\
    \ \\
    \caption{\textbf{Display Type of Statements in the ERRs.} The Venn diagram shows how information is conveyed in ERRs. About half of the statements can be made in either textual or tabular form. 980 statements only appear in text form, and 764 statements always appear in tables. Only relatively few statements, mostly related to stock price history and other market data, are only displayed in graphical form.}
    \begin{center}
        \includegraphics[width=4.0cm]{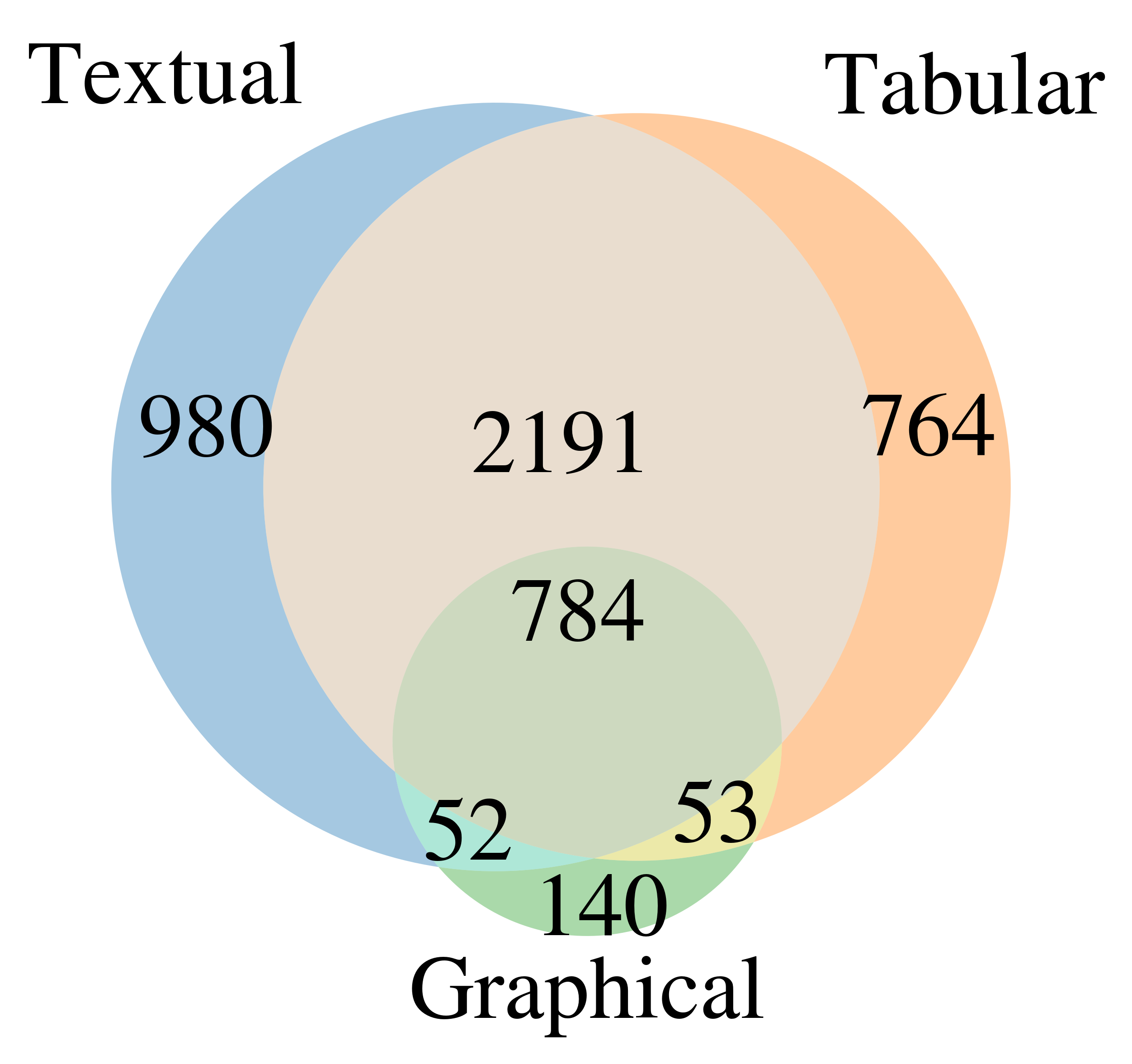}
    \end{center}\label{fig:vennDiagramModalitiesInEquityResearchReports}    
\end{figure}

\subsection{Question List and Question Categories}

The annotation process was bias-free, without presumptions about the space of questions we would encounter. Sentence-by-sentence, we read the ERRs and annotated each sentence with a question. When we encountered an answer to a question that we previously saw, we mapped the statement to the existing question. When a question was not on the list, we added a new question.

We grouped the resulting 169 questions into five categories. The categories are: \textit{Financials, Company, Product, Stock, Market, Analysis}. 

\subsection{Question Classification}

To make use of this question list for our purpose of analyzing the automation potential of ERRs, we classified each unique question in two dimensions: \begin{enumerate}
    \item \textbf{Extractability:} If the answer to a question is found in a corporate report, the question is extractable. To validate that our classification of extractability is correct, we ran the open-source Llama-3-70B and the closed-source GPT-4-turbo-2024-04-09 models on 200 example questions, showing that these language models can indeed extract the answers to the questions when provided with annual reports as the prompt context. We describe this validation step in section~\ref{subsection_validation_of_automation_potential_by_checking_if_models_can_extract}. 
    \item \textbf{Display Modality:} Refers to how analysts display the statement. The results are reported in figure~\ref{fig:vennDiagramModalitiesInEquityResearchReports},  showing that most information can be displayed in text or tabular form.
\end{enumerate}

The aggregated results are in table~\ref{tab:overview_of_extractable_and_non_extractable_counts}. 
We used company-issued reports to check whether the questions are extractable.

The annotation process required two iterations of reading through the reports. In the first reading process, the question list was created. In the second iteration, we labeled the extractability and the modality columns. The second iteration involved reading the ERRs again, and finding the source for the answer to each question. If a direct answer was matched in a single text source, the extractability was marked as ``extractable.'' If the answer was found in a financial markets database (such as Bloomberg), the question was marked as ``database-extractable.'' If not found in single text passage or database, the extractability was marked as ``non-extractable.''

\begin{figure}

    \caption{\textbf{Frequency of Different Modes of Data Representation in ERRs.} The two categories ``Tabular or Graphical Data'' (53 occurrences) and ``Textual or Graphical Data'' (52) were filtered out as only very few questions are represented in these ways. }
    \begin{center}
        \includegraphics[width=6.5cm]{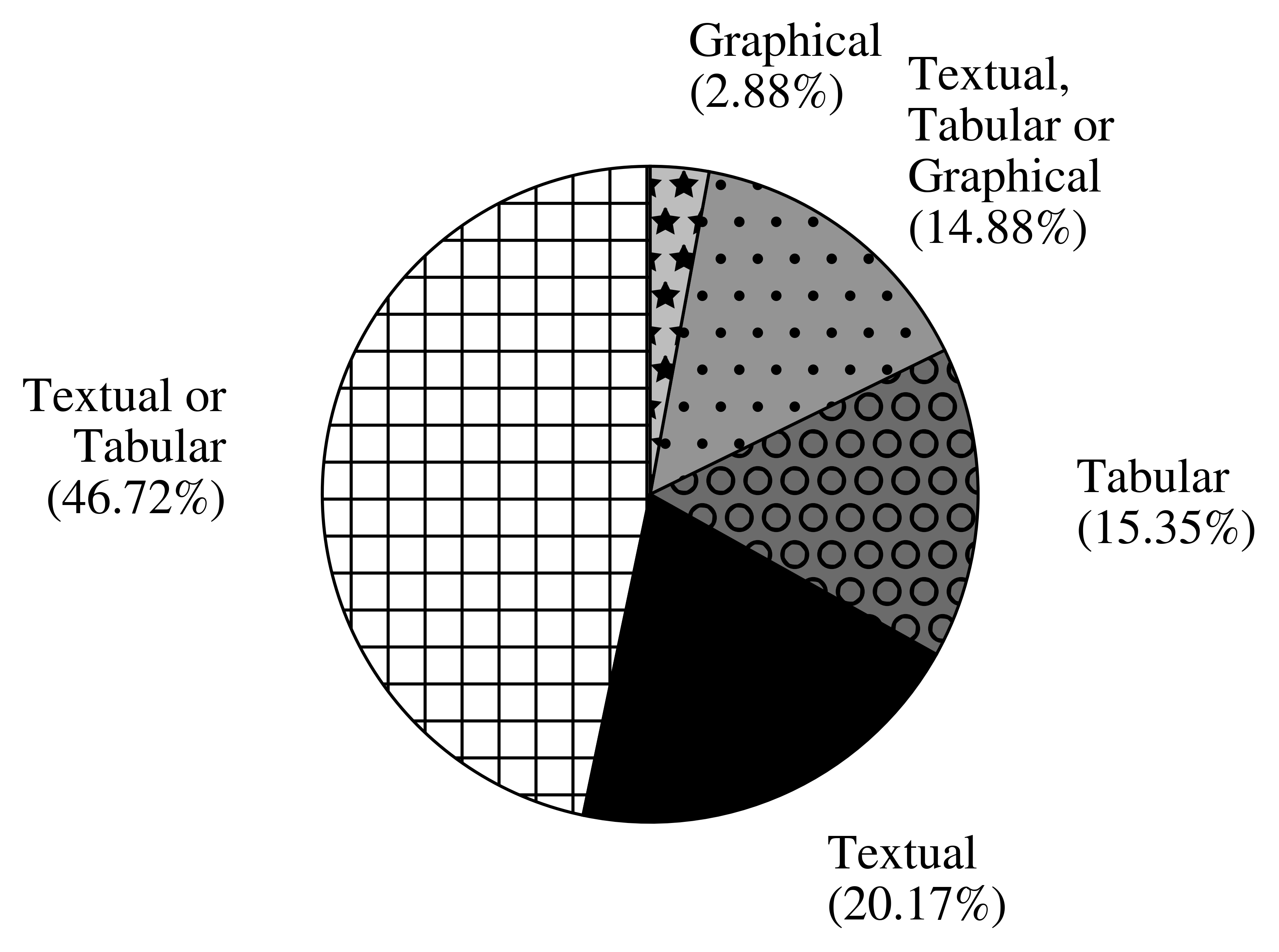}
    \end{center}\label{fig:modalities_types_of_display_as_pie_chart}
\end{figure}

\subsection{Qualitative Validation of the Question List Using Expert Interviews and Prior Research}\label{subsection_methods_qualitative_validation}

As a sanity check, we conducted designated validation 45-minute interviews with ten financial analysts. 
In the interviews, we were reassured that our results align with what would be expected in practice when considering the most important questions in ERRs.
In some interviews, however, it became clear that financial analysts consider the management qualifications and subjective impressions about the competence of managers. This aspect is not captured in this study and will be hard to capture for automated systems.


In addition to validation through qualitative interviews, prior research by \cite[p.\ 246]{AsquithMikhailAu_InformationContentOfEquityAnalystReports_2005} confirms that the top questions identified in our study match their findings about the most frequent ERR contents.

\subsection{Validation of the Automation Potential by Comparing Human, Llama-3-70B, and GPT-4-turbo-2024-04-09 Report Generation Performance}\label{subsection_validation_of_automation_potential_by_checking_if_models_can_extract}

We validate the claims made in this paper about the automation potential of specific questions by automating those parts of the report generation that we have classified as ``text-extractable.'' We test which questions are the hardest questions for language models to answer, which informs our assessment of the question ``text-extractability.''

We use the open-source Llama-3-70B model by \cite{metaAi_llama3_dot_1__2024} and the closed-source GPT-4-turbo-2024-04-09 model by \cite{achiamOpenai_gpt4_2023}. We set the temperatures of both models to zero (giving the models the chance to always use their true best guesses) and do not limit the number of output tokens. In cases where the context lengths of the models were exceeded, we split the contexts and concatenated the  outputs.

\begin{figure}
    \begin{center}
        \caption{\textbf{Share of Correct, Incorrect, Database-Extractable, and Non-Extractable Questions for GPT-4-turbo-2024-04-09 and Llama-3-70B, Weighted by Occurence.} Stock-related questions can only be answered using financial market data. Those thus fall under a separate category that is automatable, yet not by using language models. Llama~3 has a slight edge over GPT-4 as it answered more questions correctly than GPT-4 did.}

        \includegraphics[width=9cm]{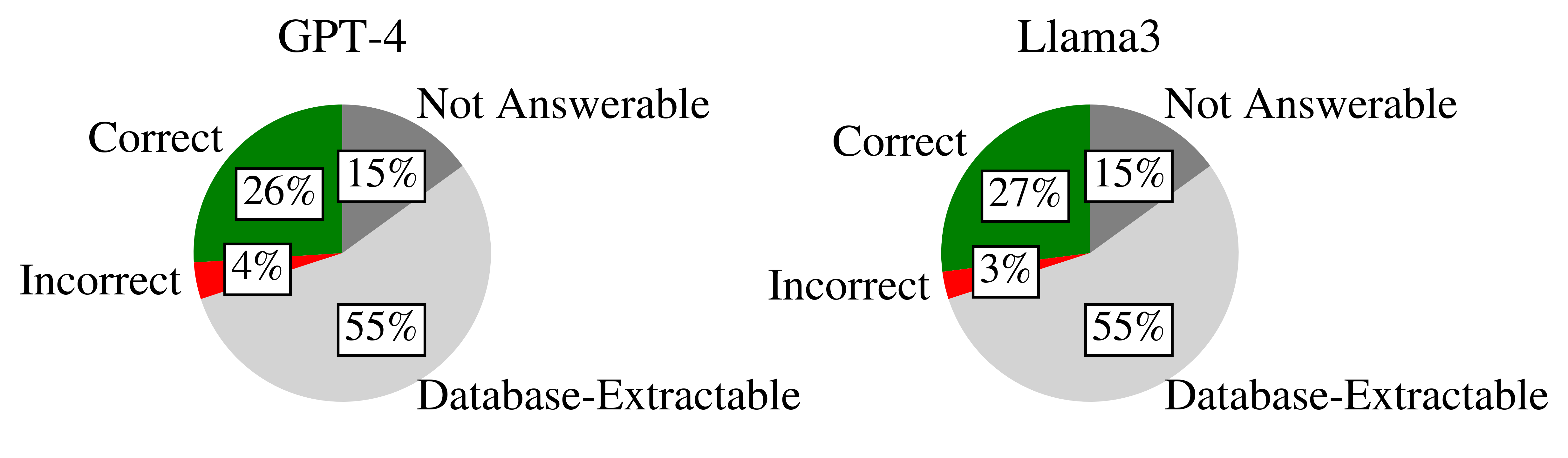}\label{fig:evaluation_gpt4_llama3__pie_charts_correct_incorrect}
    \end{center}
\end{figure}

Our results show that Llama-3-70B is able to extract information from annual reports for 27\% of the 200 questions. \mbox{GPT-4} is able to extract the correct answer in 26\% of the questions (figure~\ref{fig:evaluation_gpt4_llama3__pie_charts_correct_incorrect}). The results are in line with our expectations and confirm that the models can indeed extract the answers to the questions when provided with annual reports as the prompt context.

\begin{figure}
    \ \\
    \begin{center}
     
        \caption{\textbf{Correctness of Answers by GPT-4-turbo-2024-04-09, Llama-3-70B, and the Best of Both Models.} The green parts show correct answers, the red-white-hatched parts show errors. Interestingly, the errors of GPT-4 and Llama~3 have almost no overlap. When one model is unable to correctly answer a question, the other model usually is. There was only one question that both models did not answer correctly despite the relevant information being present in the prompt. The fine black lines in the second plot delimit different questions, and the culmination of errors for certain questions shows that the models have difficulty with particular questions.}
        \includegraphics[width=8.3cm]{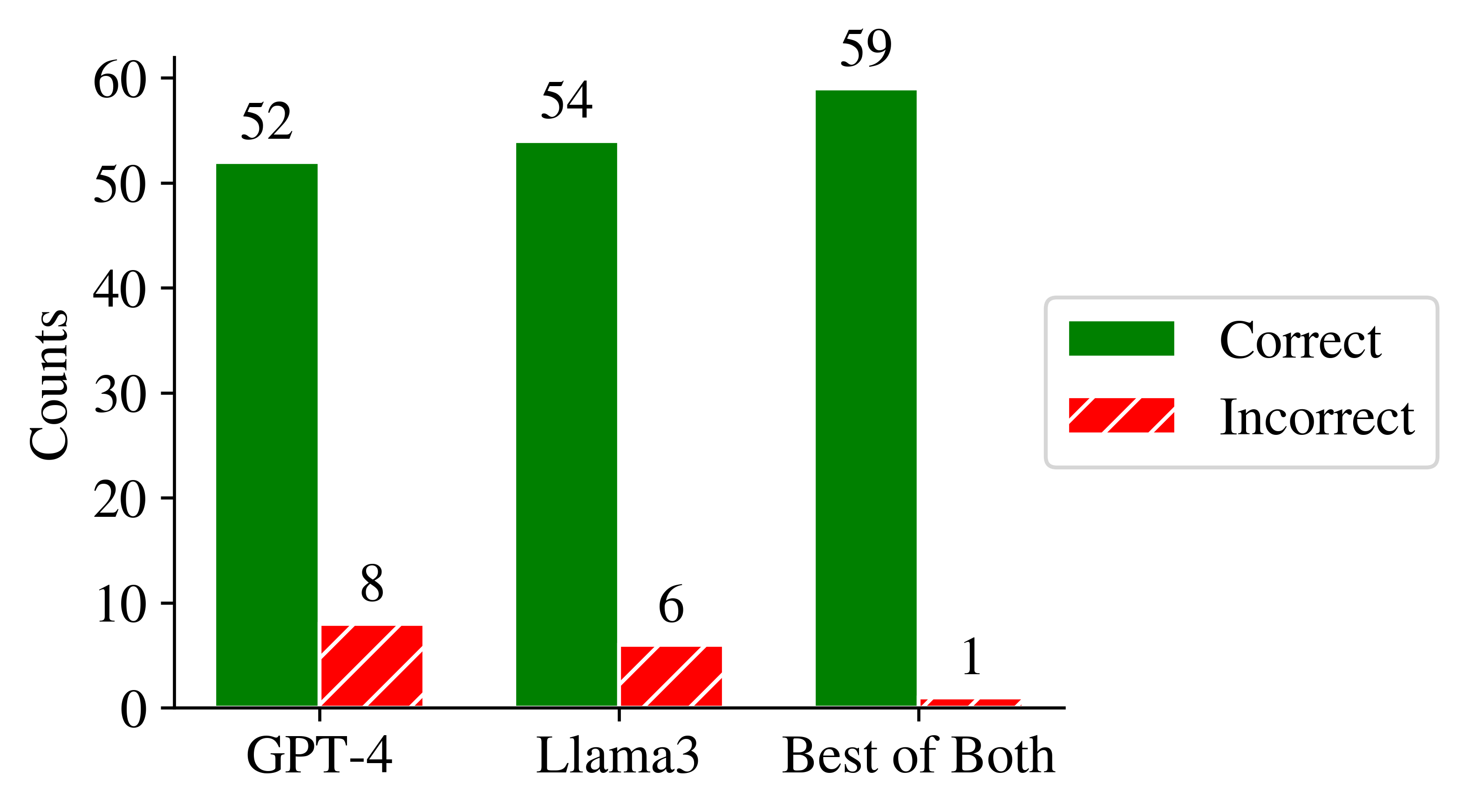}\label{fig:evaluation_gpt4_llama3__bar_charts_correct_incorrect_counts_with_either_model_and_color}
        \includegraphics[width=3.8cm]{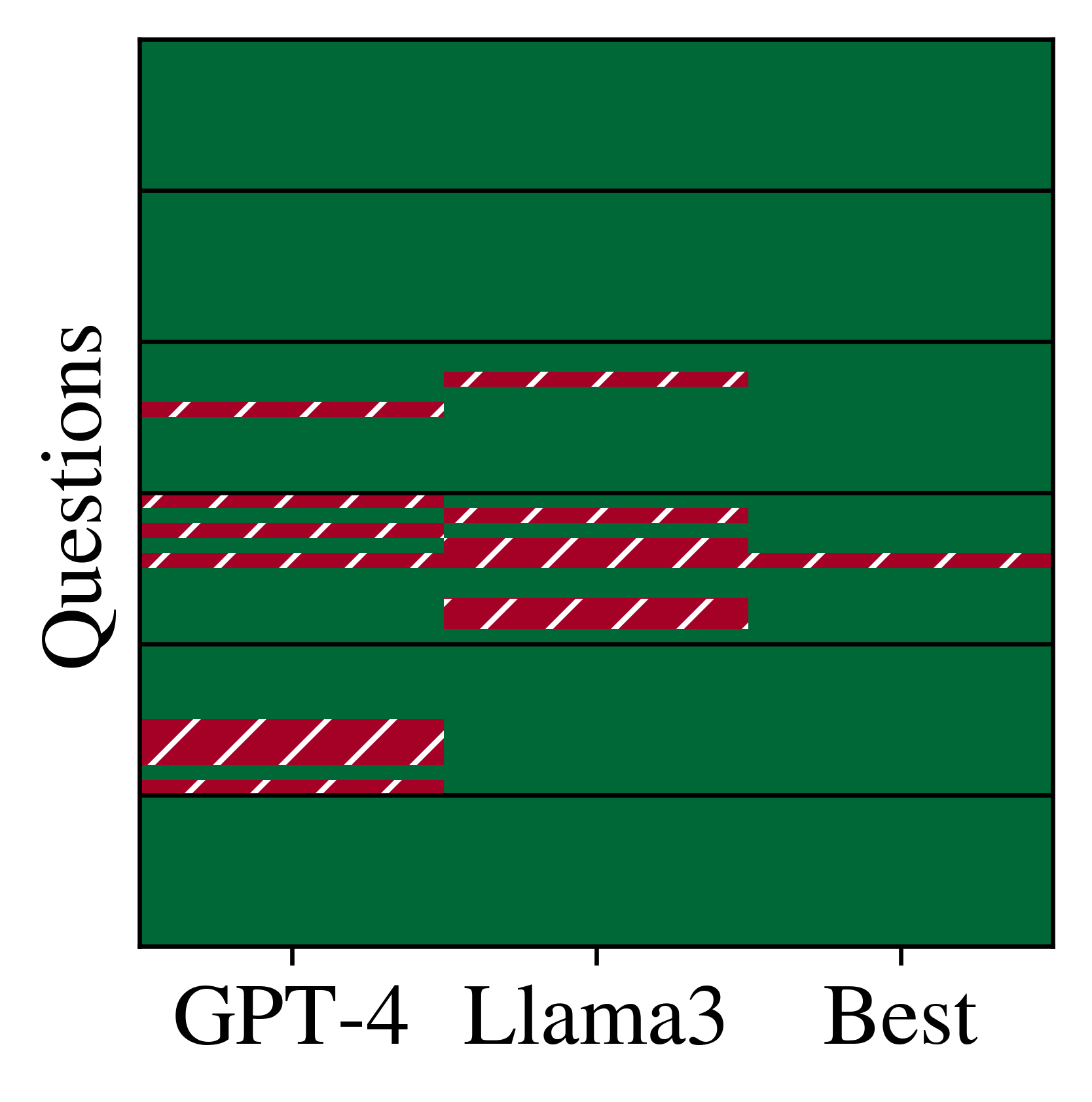}\label{fig:evaluation_gpt4_llama3__grid_chart_correctness_when_both_models_used_in_conjunction}
    \end{center}
\end{figure}

If one adds the database-extractable questions, which can be gathered automatically from financial data providers, the share of automatable questions rises to 55\% + 26\% = 81\% for \mbox{GPT-4}, and 82\% for Llama~3. If one then also considers that the models' performance is highly uncorrelated, one could use both models at once to achieve an ensemble that can answer 84\% of questions, and makes mistakes only for about 1\% of questions (see also figure \ref{fig:evaluation_gpt4_llama3__grid_chart_correctness_when_both_models_used_in_conjunction}).

As a qualitative side note, we found that GPT-4 tends to provide longer responses with more context. In some cases, we thus found that GPT-4 provided helpful context that Llama-3-70B missed. We performed some follow-up tests with Llama~3 to see if this difference was a lack of capability or simply a difference in the default verbosity among the models. We found that, for the purpose of financial text, Llama~3 is able to provide the same extraction depth and abstraction capabilities as GPT-4 does. But Llama~3 tends to provide more direct answers compared to GPT-4 unless prompted to add contextual flavor. Also, Llama often attempts to calculate growth numbers when asked about rates. In all attempts, it fails to provide correct absolute or relative year-on-year changes but stays in the correct ballpark of plus/minus 10\%.

Llama~3 was trained on sequences of 8,192 tokens. Annual reports are usually much longer, often having around 100,000 tokens. 

The language models correctly identified information in full-form text and in tabular format. Tables were simply copy-pasted from the annual reports, so the formatting of these tables was not specifically optimized for language model readability. The models showed high robustness in extracting financial information from tables. For each mistake made by the language models, we inspected the context to see if a human had been able to answer the question given the text-only context (no PDF formatting was provided, limiting what the language models were able to parse relative to what a human would be able to visually infer from the format in the annual report). We made sure that no language model answer was marked as incorrect if there was no clear answer in the context, but no such cases occurred in the sample of 200 questions.

\section{Results}\label{section_results_ffaq}

\begin{figure}
    \caption{\textbf{Share of Question Subcategories.} Please note that there are three ``Other'' labels on the x-axis. These refer to the ``Other'' subcategory of their respective categories: ``Financial -- Other,'' ``Company -- Other,'' and ``Analysis -- Other.''}
    \begin{center}
        \includegraphics[width=7.3cm]{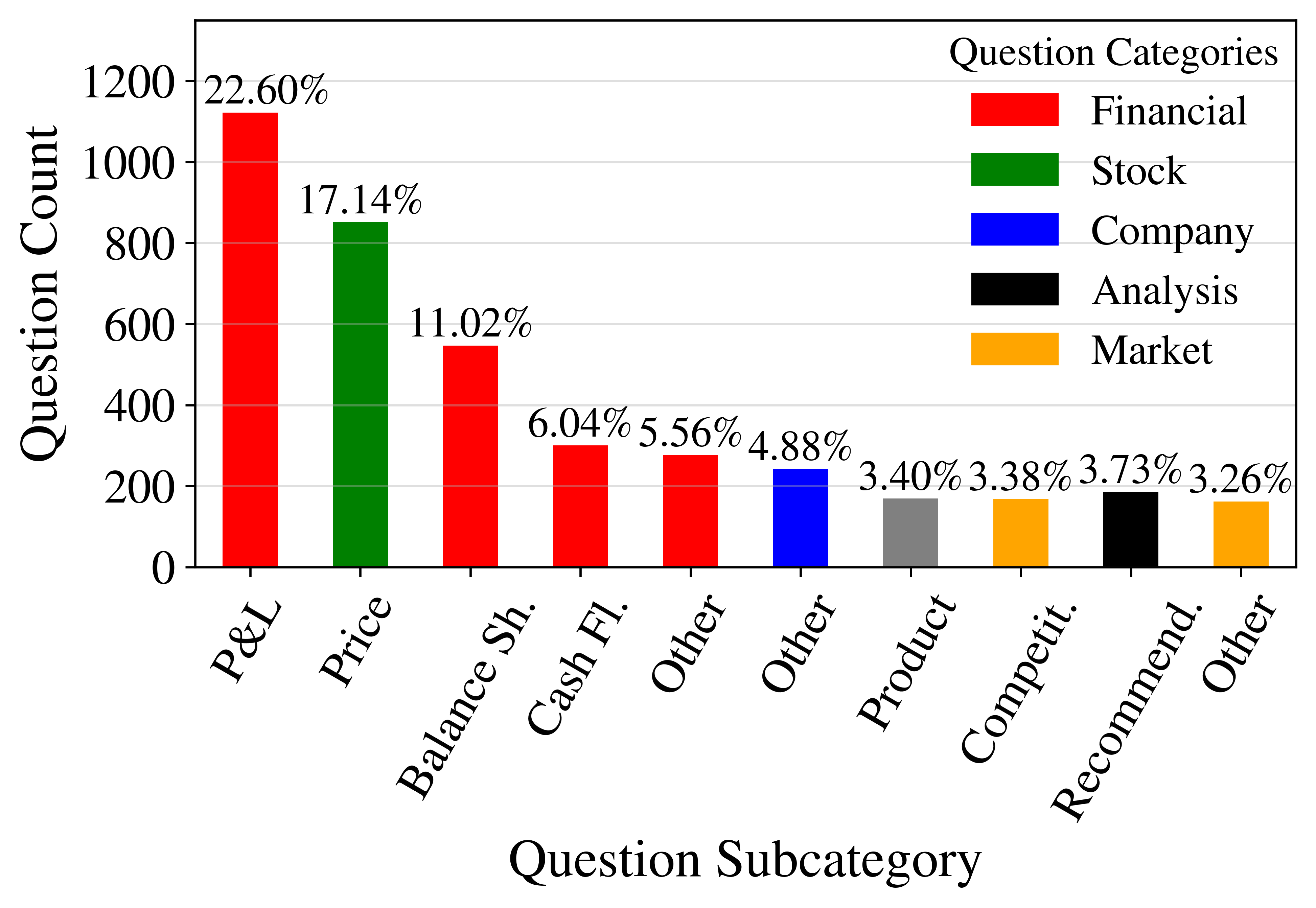}\label{fig:share_of_question_categories_x_subcategory_y_occurrences}
    \end{center}
\end{figure}

\subsection{Result Overview}

In summary, 75.15\% of the 169 questions in ERRs are automatable. More precisely, 51.91\% of the statements in ERRs are extractable, and 24.24\% of questions require access to non-public databases but have potential for automation. Only 24.85\% of questions require judgment that goes beyond extraction from either a corporate report or from a financial database.

\subsection{Analysis of Question Categories and Subcategories}

\begin{table}
    \caption{\textbf{Question Frequency List of the Five Most Frequently Answered Questions.} The table indicates which types of statements are the most frequent across all ERRs. Across the 72 reports, 66 contained information about the stock price and 65 about challenges. Company details and market environment statements appear less frequently.}
    
    \begin{center}
        {\fontsize{8}{9}\selectfont
        \addtolength{\tabcolsep}{-0.6em}
\begin{tabular}{lrlll}
    \toprule
    Question & Count & Subcategory & Numerical &  \begin{tabular}{ll}Extractable\\ From Text\end{tabular} \\
    \midrule
    Key financials & 122 & Financials - Other & Yes & Yes \\
    Analyst rating & 64 & Analysis - Recommend. & No & No \\
    Cash flow & 64 & Financials - P\&L & Yes & Yes \\
    Target price? & 62 & Analysis - Recommend. & Yes & No \\
    Revenue over time & 60 & Financials - P\&L & Yes & Yes \\
    \bottomrule
\end{tabular}
\label{table_question_frequency_list_top_5}
    } 
    \end{center}
\end{table}

A share of 73.4\% of statements in the category \textit{Product} are automatable. 
\textit{Financials} is the most critical question category by statement count. 70.6\% of statements from this category are extractable. 
A share of 54.6\% is automatable in the category \textit{Company}. 
A share of 16.6\% is automatable in the category \textit{Stock}. 
A share of 4.6\% is automatable in the category \textit{Market}. This is because statements about the market environment usually require access to diverse sources outside the company's annual and quarterly reports. 
None of the statements from the \textit{Analysis} category can be automated with extractions from publicly available corporate reports. This category contains the target price (forward guidance), recommendation, and risk assessment.

\subsection{Analysis of Extractable and Non-Extractable Statements}

\begin{table}
    \caption{\textbf{Overview of Extractable and Non-extractable Statements.} The table portrays the classifications of questions answered in the examined ERRs. The columns show whether the questions answered are numeric or non-numeric. The rows indicate whether the information is extractable from textual sources (such as annual reports).}

    \begin{center}
        Counted by the number of unique questions:

        \begin{tabular}{lll}
\toprule
 & Numeric & Non-Numeric \\
\midrule
Extractable From Text & 67 (40.61\%) & 17 (10.3\%) \\
Not Extractable & 42 (25.45\%) & 39 (23.64\%) \\
\bottomrule
\end{tabular}

        \ \\ 

        \ \\ 

        Counted by the total number of statement occurrences:

        \begin{tabular}{lll}
\toprule
 & Numeric & Non-Numeric \\
\midrule
Extractable From Text & 1925 (38.78\%) & 437 (8.8\%) \\
Not Extractable & 1425 (28.71\%) & 1177 (23.71\%) \\
\bottomrule
\end{tabular}

    \end{center}\label{tab:overview_of_extractable_and_non_extractable_counts}
\end{table}

The first part of table~\ref{tab:overview_of_extractable_and_non_extractable_counts} contains the classification of the 165 unique questions from all ERRs. Two-thirds of the questions are numeric, and more than half are extractable. While extractable information is mostly numeric (40.61\% of total questions are numeric-extractable, 61.47\% of numeric questions are extractable), extractable non-numeric information is rare (10.3\% of total questions, 30.36\% of non-numeric questions are extractable). Out of the non-extractable information, slightly more is numeric, but the number of numeric questions is higher (109 unique questions) than the number of non-numeric questions (56 unique questions).

The \textit{Analysis} category requires special mention as it contains summarizing statements that make recommendations. These statements are not extractable from anywhere, as they require comprehension across multiple sources. Only 3.64\% of statements (non-unique) in ERRs fall under the analysis category. 

Table~\ref{tab:overview_of_extractable_and_non_extractable_counts} shows that, without weighting the questions by their occurrence frequency, 50.91\% (40.61\% + 10.3\%) of questions answered in ERRs can be answered by extracting information from public textual sources. 





\subsection{Analysis of Contextualizing and Summarizing Components}\label{subsection_results_analysis_of_contextualizing_and_summarizing_components}

The \textit{Analysis} question category contains summarizing and contextualizing components. 
Given the same set of facts, different analysts may weigh, select, and combine those facts differently, leading to different recommendations.

Related to this, there are numerous questions in the category \textit{Market} of similar nature. Market developments require simplification and curation to distill into a few pages of text. Similar to the \textit{Analysis} category, different observers judge the same set of facts differently, leading to different conclusions. Given that the potential inputs to this category are vast, with many news reports and other sources to choose from, it is unlikely that an automated system can already handle this task.

\subsection{Tabular Data}

Most statements in ERRs are textual or tabular. \cite{ChenChenSmileyShahBorovaLangdonMoussaBeaneHuangRoutledgeWang_FinQaADatasetOfNumericalReasoningOverFinancialData_2021} show that numerical reasoning across tables and text is feasible (pp.\ 5--7). \cite{zhuLeiHuangWangZhangLvFengChua_TatQa_AQuestionAnsweringBenchmarkOnAHybridOfTabularAndTextualContentInFinance_2021} confirm that tabular information extraction is possible, particularly for financial data (pp.\ 3282--3284).

We confirm these findings: Our validation from section~\ref{subsection_validation_of_automation_potential_by_checking_if_models_can_extract} has not required any manual formatting of table data -- we copied tables from annual reports without formatting into the models' context, and they extracted information from these ill-formatted strings with high reliability.

\section{Conclusion}\label{section_conclusion_ffaq}

\begin{table}
    \caption{\textbf{List of Research Providers, Sorted by the Number of Annual Reports Used in This Study.} J.P.\ Morgan provided the most equity research reports for this analysis with 16 pieces, followed by Deutsche Bank (9), Zacks (8), and Barclays (7). The average number of statements is 68.6, the median is 70, and the minimum number is a result of Veritas Investment Research's one-pager with only nine statements.}
    \begin{center}\label{tab:AnalystHouses_Research_Providers_Banks}
        {\fontsize{8}{9}\selectfont
\begin{tabular}{|l|c|c|} \hline 
    Research Provider   & \begin{tabular}{@{}c@{}} Research Report \\ Counts\end{tabular}   & \begin{tabular}{@{}c@{}} Avg.\ No.\ Statements \\ per Report\end{tabular}\\ 

    \hline \hline  

    J.P.\ Morgan    & 16                        & 93\\ \hline  
    Deutsche Bank    & 9                        & 56\\ \hline 
    Zacks    & 8                        & 90\\ \hline 
    Barclays    & 7                        & 63\\ \hline 
    Mizuho    & 5                        & 48\\ \hline 
    Needham    & 5                        & 71\\ \hline 
    KBW    & 3                       & 55\\ \hline 
    Refinitiv    & 2                       & 53\\ \hline 
    New Constructs    & 2                       & 40\\ \hline  
    Phillip Securities Res.   & 2                       & 78\\ \hline  
    GlobalData    & 1                       & 115\\ \hline 
    China Renaissance    & 1                       & 106\\ \hline
    IBM Res. &	1	&34\\ \hline 
    Punto Casa de Bolsa &	1 &	24\\ \hline 
    Spartan Capital &	1 &	58\\ \hline 
    Thompson Res.	 &1  &	34\\ \hline 
    Mitsubishi UFJ M.S.	 &1	 &31\\ \hline 
    Oppenheimer	 &1	 &44\\ \hline 
    BPC Res.	 &1	 &44\\ \hline 
    Veritas Investment &	1 &	9\\ \hline 
    Echelon	 & 1	 & 59\\ \hline 
    finnCap	 & 1	 & 75\\ \hline 
    BTIG	 & 1	 & 69\\ \hline 
\end{tabular}
}
    \end{center}

\end{table}

\subsection{Summary}

Our results confirm the findings by~\cite{colemanMerkleyPacelli_HumanVersusMachineAComparisonOfRoboAnalystAndTraditionalResearchAnalystInvestmentRecommendations_2022} that partly automating equity research reports (ERRs) is feasible. Only one quarter of questions require complex judgment that takes into consideration more information than would fit in a language model's context window.

Given the oversized importance of the \textit{Analysis} category, and given that humans may still be better at providing high-stake recommendations, this category may be hard to automate with current models and be left to human financial analysts. It constitutes 3.64\% of all statements of ERRs.

Another finding is that model errors often do not overlap (figure~\ref{fig:evaluation_gpt4_llama3__grid_chart_correctness_when_both_models_used_in_conjunction}). Language models for information extraction show promising performance for extracting financial data. As this data is relevant to ERRs, the partial automation of ERRs appears feasible, especially when ensemble models are used that have independent blind spots.

\subsection{Limitations}\label{subsection_limitations_ffaq}


Our counting approach does not weigh the importance of the questions. The most important questions may appear less frequently. Furthermore, there could be out-of-distribution questions that we did not capture in this analysis because they were not present in the ERRs we analyzed.\label{limitations_out_of_distribution_questions_specialized_ffaq}



Only 72 ERRs from 23 research firms were dissected. Other research firms may include questions not in the space of 169 question archetypes identified in this study. 






\subsection{Future Work}\label{subsection_future_work_ffaq}

Direct information extraction for ERRs is still a largely unexplored field. Various technical methods were presented in the literature section~\ref{section_literature_review_automation_of_equity_research_ffaq}. Future research can implement these methods to write ERRs automatically.

Future research can produce benchmarks for ERR generation and add those to existing language model evaluation suites, adding to prior work \cite{MaloSinhaKorhonenWallenius_FinancialPhraseBank_GoodDebtOrBadDebtDetectingSemanticOrientationsinEconomicTexts_2014,maiaHandschuhFreitagDavidMcDermott_2018_WWWOpenChallengeFinancialOpinionMiningAndQuestionAnswering,zhuLeiHuangWangZhangLvFengChua_TatQa_AQuestionAnsweringBenchmarkOnAHybridOfTabularAndTextualContentInFinance_2021, ChenChenSmileyShahBorovaLangdonMoussaBeaneHuangRoutledgeWang_FinQaADatasetOfNumericalReasoningOverFinancialData_2021} by including very long contexts with raw annual report text.

In addition to creating suitable benchmarks, future research can develop domain-configured models that can generate ERRs from realistic sources for financial information (such as annual reports and quarterly reports). Human evaluators or standardized benchmarks can access the performance of such models.

\bibliographystyle{IEEEtran}

\bibliography{references}

\begin{thebibliography}{10}
\providecommand{\url}[1]{#1}
\csname url@samestyle\endcsname
\providecommand{\newblock}{\relax}
\providecommand{\bibinfo}[2]{#2}
\providecommand{\BIBentrySTDinterwordspacing}{\spaceskip=0pt\relax}
\providecommand{\BIBentryALTinterwordstretchfactor}{4}
\providecommand{\BIBentryALTinterwordspacing}{\spaceskip=\fontdimen2\font plus
\BIBentryALTinterwordstretchfactor\fontdimen3\font minus \fontdimen4\font\relax}
\providecommand{\BIBforeignlanguage}[2]{{%
\expandafter\ifx\csname l@#1\endcsname\relax
\typeout{** WARNING: IEEEtran.bst: No hyphenation pattern has been}%
\typeout{** loaded for the language `#1'. Using the pattern for}%
\typeout{** the default language instead.}%
\else
\language=\csname l@#1\endcsname
\fi
#2}}
\providecommand{\BIBdecl}{\relax}
\BIBdecl

\bibitem{AsquithMikhailAu_InformationContentOfEquityAnalystReports_2005}
P.~Asquith, M.~Mikhail, and A.~Au, ``Information content of equity analyst reports,'' \emph{Journal of Financial Economics}, vol.~75, no.~2, pp. 245--282, 2005.

\bibitem{colemanMerkleyPacelli_HumanVersusMachineAComparisonOfRoboAnalystAndTraditionalResearchAnalystInvestmentRecommendations_2022}
B.~Coleman, K.~Merkley, and J.~Pacelli, ``Human versus machine: A comparison of robo-analyst and traditional research analyst investment recommendations,'' \emph{The Accounting Review}, vol.~97, no.~5, pp. 221--244, 2022.

\bibitem{bjerringLakonishokVermaelen_StockPricesAndFinancialAnalystsRecommendations_1983}
J.~Bjerring, J.~Lakonishok, and T.~Vermaelen, ``Stock prices and financial analysts' recommendations,'' \emph{The Journal of Finance}, vol.~38, no.~1, pp. 187--204, 1983.

\bibitem{elton_GruberGrossman_DiscreteExectationalDataAndPortfolioPerformance_1986}
E.~Elton, M.~Gruber, and S.~Grossman, ``Discrete expectational data and portfolio performance,'' \emph{The Journal of Finance}, vol.~41, no.~3, pp. 699--713, 1986.

\bibitem{liu1_SmithSyed_StockPriceReactionsToTheWallStreetJournalsSecuritiesRecommendations_1990}
P.~Liu, S.~Smith, and A.~Syed, ``Stock price reactions to the {Wall Street Journal}'s securities recommendations,'' \emph{Journal of Financial and Quantitative Analysis}, vol.~25, no.~3, pp. 399--410, 1990.

\bibitem{beneish_StockPricesAndTheDisseminationOfAnalystsRecommendation_1991}
M.~Beneish, ``Stock prices and the dissemination of analysts' recommendation,'' \emph{Journal of Business}, pp. 393--416, 1991.

\bibitem{stickel_TheAnatomyOfThePerformanceOfBuyAndSellRecommendations_1995}
S.~Stickel, ``The anatomy of the performance of buy and sell recommendations,'' \emph{Financial Analysts Journal}, pp. 25--39, 1995.

\bibitem{Womack_DoBrokerageAnalystsRecommendationsHaveInvestmentValue_1996}
K.~Womack, ``Do brokerage analysts' recommendations have investment value?'' \emph{The Journal of Finance}, vol.~51, no.~1, pp. 137--167, 1996.

\bibitem{barberLehavyMcnicholsTrueman_CanInvestorsProfitFromTheProphetsSecurityAnalystRecommendationsAndStockReturns_2001}
B.~Barber, R.~Lehavy, M.~McNichols, and B.~Trueman, ``Can investors profit from the prophets? {Security} analyst recommendations and stock returns,'' \emph{The Journal of Finance}, vol.~56, no.~2, pp. 531--563, 2001.

\bibitem{MikhailWaltherWillis_DoSecurityAnalystsExhibitPersistentDifferencesInStockPickingAbility_2004}
M.~Mikhail, B.~Walther, and R.~Willis, ``Do security analysts exhibit persistent differences in stock picking ability?'' \emph{Journal of Financial Economics}, vol.~74, no.~1, pp. 67--91, 2004.

\bibitem{michaelyWomack_ConflictOfInterestAndTheCredibilityOfUnderwriterAnalystRecommendations_1999}
R.~Michaely and K.~Womack, ``Conflict of interest and the credibility of underwriter analyst recommendations,'' \emph{The Review of Financial Studies}, vol.~12, no.~4, pp. 653--686, 1999.

\bibitem{BradshawBrownHuang_DoSellSideAnalystsExhibitDifferentialTargetPriceForecastingAbility_2013}
M.~Bradshaw, L.~Brown, and K.~Huang, ``Do sell-side analysts exhibit differential target price forecasting ability?'' \emph{Review of Accounting Studies}, vol.~18, no.~4, pp. 930--955, 2013.

\bibitem{BoniniZanettiBianchiniSalvi_TargetPriceAccuracyInEquityResearch_2010}
S.~Bonini, L.~Zanetti, R.~Bianchini, and A.~Salvi, ``Target price accuracy in equity research,'' \emph{Journal of Business Finance \& Accounting}, vol.~37, no. 9-10, pp. 1177--1217, 2010.

\bibitem{GleasonJohnsonLi_ValuationModelUseAndThePriceTargetPerformanceOfSellSideEquityAnalysts_2013}
C.~Gleason, B.~Johnson, and H.~Li, ``Valuation model use and the price target performance of sell-side equity analysts,'' \emph{Contemporary Accounting Research}, vol.~30, no.~1, pp. 80--115, 2013.

\bibitem{FamaFrench_TheCrossSectionOfExpectedStockReturns_1992}
E.~Fama and K.~French, ``The cross-section of expected stock returns,'' \emph{The Journal of Finance}, vol.~47, no.~2, pp. 427--465, 1992.

\bibitem{FamaFrench_SizeAndBookToMarketFactorsInEarningsAndReturns_1995}
------, ``Size and book-to-market factors in earnings and returns,'' \emph{The Journal of Finance}, vol.~50, no.~1, pp. 131--155, 1995.

\bibitem{FamaFrench_ChoosingFactors_2018}
------, ``Choosing factors,'' \emph{Journal of Financial Economics}, vol. 128, no.~2, pp. 234--252, 2018.

\bibitem{FamaFrench_AFiveFactorAssetPricingModel_2015}
------, ``A five-factor asset pricing model,'' \emph{Journal of Financial Economics}, vol. 116, no.~1, pp. 1--22, 2015.

\bibitem{Carhart_OnPersistenceInMutualFundPerformance_1997}
M.~Carhart, ``On persistence in mutual fund performance,'' \emph{The Journal of Finance}, vol.~52, no.~1, pp. 57--82, 1997.

\bibitem{Asness_ValueAndMomentumEverywhere_2013}
C.~Asness, T.~Moskowitz, and L.~Pedersen, ``Value and momentum everywhere,'' \emph{The Journal of Finance}, vol.~68, no.~3, pp. 929--985, 2013.

\bibitem{kimMuhnNikolaev_FinancialStatementAnalysisWithLargeLanguageModels_2024}
A.~Kim, M.~Muhn, and V.~Nikolaev, ``Financial statement analysis with large language models,'' \emph{Chicago Booth Research Paper Forthcoming, Fama-Miller Working Paper}, 2024.

\bibitem{ChenChenSmileyShahBorovaLangdonMoussaBeaneHuangRoutledgeWang_FinQaADatasetOfNumericalReasoningOverFinancialData_2021}
Z.~Chen, W.~Chen, C.~Smiley, S.~Shah, I.~Borova, D.~Langdon, R.~Moussa, M.~Beane, T.-H. Huang, B.~Routledge, and W.~Y. Wang, ``{F}in{QA}: A dataset of numerical reasoning over financial data,'' \emph{Conference on Empirical Methods in Natural Language Processing (EMNLP)}, pp. 3697--3711, 2021.

\bibitem{thaiDavidORiainOSullivanHandschuh_SemanticallyEnhancedPassageRetrievalForBusinessAnalysisActivity_2008}
V.~Thai, B.~Davis, S.~O'Riain, D.~O'Sullivan, and S.~Handschuh, ``Semantically enhanced passage retrieval for business analysis activity,'' \emph{European Conference on Information Systems (ECIS)}, 2008.

\bibitem{BorgeaudMenschHoffmann_DeepMind_RETRO_ImprovingLanguageModelsByRetrievingFromTrillionsOfTokens_2021}
S.~Borgeaud, A.~Mensch, J.~Hoffmann, T.~Cai, E.~Rutherford, K.~Millican \emph{et~al.}, ``Improving language models by retrieving from trillions of tokens,'' \emph{International Conference on Machine Learning, Proceedings of Machine Learning Research (PMLR)}, vol. 162, 2021.

\bibitem{chenWongChenTian_ExtendingContextWindowOfLargeLanguageModelsViaPositionalInterpolation_2023}
S.~Chen, S.~Wong, L.~Chen, and Y.~Tian, ``Extending context window of large language models via positional interpolation,'' \emph{arXiv}, 2023.

\bibitem{karpukhinOguzMinLewisWuEdunovChenYih_DensePassageRetrievalForOpenDomainQuestionAnsewring_2020}
V.~Karpukhin, B.~Oguz, S.~Min, P.~Lewis, L.~Wu, S.~Edunov, D.~Chen, and W.-t. Yih, ``Dense passage retrieval for open-domain question answering,'' \emph{Conference on Empirical Methods in Natural Language Processing (EMNLP)}, pp. 6769--6781, 2020.

\bibitem{LewisPerezPiktusPetroniGoyalKuttlerLewisYihRocktaschelRiedelKiela_RetrievalAugmentedGenerationForKnowledgeIntensiveNlpTasks_2020}
P.~Lewis, E.~Perez, A.~Piktus, F.~Petroni, V.~Karpukhin, N.~Goyal, H.~K\"{u}ttler, M.~Lewis, W.-t. Yih, T.~Rockt\"{a}schel, S.~Riedel, and D.~Kiela, ``Retrieval-augmented generation for knowledge-intensive nlp tasks,'' \emph{Advances in Neural Information Processing Systems}, vol.~33, pp. 9459--9474, 2020.

\bibitem{GuuLeeTungPasupatChang_REALM_RetrievalAugmentedLanguageModelPretraining_2020}
K.~Guu, K.~Lee, Z.~Tung, P.~Pasupat, and M.-W. Chang, ``{REALM:} {Retrieval}-augmented language model pre-training,'' \emph{International Conference on Machine Learning, PMLR}, pp. 3929--3938, 2020.

\bibitem{izacardLewisLomeliHosseiniPetroniSchickDwivediyuJoulinRiedelGrave_Atlat_FewShotLearningWithRetrievalAugmentedLanguageModels}
G.~Izacard, P.~Lewis, M.~Lomeli, L.~Hosseini, F.~Petroni, T.~Schick, J.~Dwivedi-Yu, A.~Joulin, S.~Riedel, and E.~Grave, ``Atlas: Few-shot learning with retrieval augmented language models,'' \emph{arXiv}, 2022.

\bibitem{brownMannRyderSubbiahKaplanDhariwal_GTP3_LanguageModelsAreFewShotLearners_OpenAi_2020}
T.~Brown, B.~Mann, N.~Ryder, M.~Subbiah, J.~Kaplan, P.~Dhariwal \emph{et~al.}, ``Language models are few-shot learners,'' \emph{Advances in Neural Information Processing Systems (NeurIPS)}, vol.~33, pp. 1877--1901, 2020.

\bibitem{touvronLavrilIzacardMartinetLachausLacroixRoziereGoyalHambroAzharRodriguezJoulinGraveLample_Llama_OpenAndEfficientFoundationLanguageModels_2023}
H.~Touvron, T.~Lavril, G.~Izacard, X.~Martinet, M.-A. Lachaux, T.~Lacroix, B.~Rozi{\`e}re, N.~Goyal, E.~Hambro, F.~Azhar \emph{et~al.}, ``{LLaMA}: Open and efficient foundation language models,'' \emph{arXiv}, 2023.

\bibitem{PressSmithLewis_TrainShortTestLongAttentionWithLinearBiasesEnablesInputLengthExtrapolation_2022}
O.~Press, N.~Smith, and M.~Lewis, ``Train short, test long: Attention with linear biases enables input length extrapolation,'' \emph{International Conference of Learning Representations (ICLR)}, 2022.

\bibitem{WuIsroyLuDabravolskiDredzeGehrmannKambadurRosenbergMann_BloombergGPT_ALargeLanguageModelForFinance_2023}
S.~Wu, O.~Irsoy, S.~Lu, V.~Dabravolski, M.~Dredze, S.~Gehrmann, P.~Kambadur, D.~Rosenberg, and G.~Mann, ``{BloombergGPT:} {A} large language model for finance,'' \emph{arXiv}, 2023.

\bibitem{metaAi_llama3_dot_1__2024}
Meta, ``The llama 3 herd of models,'' \emph{Technical Report}, 2024, a detailed contributor list can be found in the appendix of this paper.

\bibitem{achiamOpenai_gpt4_2023}
{OpenAI}, ``Gpt-4 technical report,'' \emph{arXiv}, 2023, the author list is excessively long with more than 200 authors and can thus be found in the technical report only.

\bibitem{zhuLeiHuangWangZhangLvFengChua_TatQa_AQuestionAnsweringBenchmarkOnAHybridOfTabularAndTextualContentInFinance_2021}
F.~Zhu, W.~Lei, Y.~Huang, C.~Wang, S.~Zhang, J.~Lv, F.~Feng, and T.-S. Chua, ``{TAT-QA:} {A} question answering benchmark on a hybrid of tabular and textual content in finance,'' in \emph{Annual Meeting of the ACL and International Joint Conference on Natural Language Processing}, 2021, pp. 3277--3287.

\bibitem{MaloSinhaKorhonenWallenius_FinancialPhraseBank_GoodDebtOrBadDebtDetectingSemanticOrientationsinEconomicTexts_2014}
P.~Malo, A.~Sinha, P.~Korhonen, J.~Wallenius, and P.~Takala, ``Good debt or bad debt: Detecting semantic orientations in economic texts,'' \emph{Journal of the Association for Information Science and Technology}, vol.~65, no.~4, pp. 782--796, 2014.

\bibitem{maiaHandschuhFreitagDavidMcDermott_2018_WWWOpenChallengeFinancialOpinionMiningAndQuestionAnswering}
M.~Maia, S.~Handschuh, A.~Freitas, B.~Davis, R.~McDermott, M.~Zarrouk, and A.~Balahur, ``{WWW'18 Open Challenge:} financial opinion mining and question answering,'' \emph{Companion Proceedings of the Web Conference}, pp. 1941--1942, 2018.

\end{thebibliography}


\end{document}